\documentclass[12pt]{iopart}

\usepackage{iopams} 
\usepackage{subfigure}   
\usepackage[graphicx]{realboxes}   
\usepackage{hyperref}  
\expandafter\let\csname equation*\endcsname\relax

\expandafter\let\csname endequation*\endcsname\relax
\usepackage{amsmath,amssymb}
\begin{document}

\title{“Classical” coherent state generated by curved surface}

\author{Weifeng Ding}
\address{ Zhejiang Province Key Laboratory of Quantum Technology and Device, School of Physics, Zhejiang University, Hangzhou 310027, China}
\ead{dingweifengtgt@163.com}
\author{Zhaoying Wang}

\address{ Zhejiang Province Key Laboratory of Quantum Technology and Device, School of Physics, Zhejiang University, Hangzhou 310027, China}
\ead{zhaoyingwang@zju.edu.cn}

\vspace{10pt}
\begin{indented}
\item[]August 2022
\end{indented}

\begin{abstract}
Analogous coherent states are deduced from classical optical fields on curved surface in this paper. The Gaussian laser beam, as a fundamental mode, cannot be adequately simulated by coherent states due to their inherent diffraction in flat space. But things will be different when it propagates on a surface with uniform curvature called the constant Gaussian curvature surface (CGCS). By generalizing the method of Feynman path integral, an equivalent coherent states solution is demonstrated here to describe the beam propagation. The temporal evolution of the Schrodinger equation is analogously translated into a spatial transmission in this derivation, we obtain the expression of quantized momentum transmitted on curved surface, which is proportional to the square root of the Gaussian curvature $K$. In addition, we build a beam propagation picture identical to the squeezed state. We hope this research can give a new view on the quantum field in curved space. 
\end{abstract}

%
%
%
%
%

\section{Introduction}

Recent experimental observations indicate that our spatial universe may be a universe with slightly negative spatial curvature \cite{RN1,RN2}. Friedmann-Robertson-Walker’s (FRW) space-time is significant as a uniform and isotropic space-time model that satisfies cosmological principles \cite{RN3}. On the other hand, the electromagnetic fields in curved spacetime exhibit a diverse set of features \cite{RN4,RN5,RN6,RN7,RN8}. This demonstrates that studying light beam propagation in a flat vacuum is insufficient in the field of optics.

Beam propagation in curved space-time is an extensive and comprehensive research field, it is proved that when the radius of curvature is comparable to the wavelength, only the intrinsic curvature of space-time affects the local propagation properties of light \cite{RN9,RN10}. Despite general relativity (GR)'s triumph, gravitational effect are too weak in a laboratory environment. One of the analog model of GR is two dimensional (2D) curved surface embedded in 3D space as a result of the 3+1 membrane paradigm theory \cite{RN30}. When taking a constant time and extracting the equatorial slice of FRW space-time, the remnant metrics can be regard as a 2D curved surface named Constant Gaussian curvature surfaces (CGCS). Since the concept was brought up by Batz and Peschel \cite{RN13}, the light propagation and electromagnetic dynamics on 2D curved surface have flourished both theoretically \cite{RN10,RN11,RN12,RN21,RN31,RN32} and experimentally \cite{RN5,RN9,RN29,RN33,RN34}. Mathematically, the different Laplacian operators introduced by curved spaces in Helmholtz equations can be further simplified by constructing spatial effective potential, which is a convenient analogy to the Schrodinger equation \cite{RN11,RN12,RN13}. Intriguingly, in this paper, we discover that the method of analogy may not only be a mathematical simplification, but also profoundly physical. We could even construct the classical harmonic oscillator potential to produce an oscillatory non-diffracted light that resembles coherent state. 

The coherent states were invented by Erwin Schrodinger in the physical context of the quantum harmonic oscillator. Lately, they were used in Quantum optics (QO) by Glauber, Sudarshan and Klauder \cite{RN14,RN15}. The anticipated time-invariant Gaussian wave packet (in quadrature representation) might be a "decent" simulation of the coherent state, although it is not fully consistent. However, the propagation of Gaussian beam in flat space will usually be diffracted. In theory, we can restrict the divergence of the light beam by setting the initial transverse mode distribution of the optical field or constructing an effective potential energy. In the experiment, the means of controlling the energy transmission of optical field are becoming increasingly abundant \cite{RN16,RN18,RN19}. The generation ways of coherent states are rich, for example by classical oscillating currents or four-wave mixing \cite{RN21}. In this paper, we consider theoretically how the bending of space itself modulates the propagation of light beam and how to originally obtain the so-called "classical" coherent states.

\section{Path integral method and quantized momentum}
In this paper, we mainly study the light paraxial transmission on a CGCS. The line element of the light on CGCS can be expressed as follows \cite{RN25}:
\begin{equation}
	\centering
	d{l^2} = d{h^2} + {\cos ^2}\left( {\sqrt K h} \right)d{\rho ^2}.
	\label{eq2}
\end{equation}
Here,$\rho$ is the equator line coordinate and $h$ is the horizontal distance from the equator. 

The Helmholtz equation when neglecting polarization effects can be written as $\left( {{\Delta _g} + {k^2}} \right)\Psi + \left( {{H^2} - K} \right)\Psi  = 0$. Here, $k = 2\pi /\lambda $. The covariant Laplacian ${\Delta _g} = \left( {1/\sqrt g } \right){\partial _i}\sqrt g {g^{ij}}{\partial _j}$, with $i,j = h,\rho$ and $g = \det \left( {{g_{ij}}} \right) = K{r^2}$, where $r={\cos }\left( {\sqrt K h} \right)/\sqrt K$ describing the generating line of the surface of rotation, i.e. the horizontal distance from the surface to the axis of rotation. The influence of the extrinsic curvature $H$ is overlooked in this paper as previous literature \cite{RN9,RN10,RN11,RN12,RN13,RN23,RN24}. If we construct the wave function $\Psi  = A{r^{ - 1/2}}u\left( {h,\rho } \right)\exp \left( {ik\rho } \right)$, $A$ is the intensity coefficient of light. Using Wenzel-Kramers-Brillouin (WKB) approximation, the Helmholtz equation can be reduced to:
\begin{equation}
	\centering
	2ik{\partial _\rho }u =  - K{r^2}{\partial ^2}_hu + {V_{eff}}u.
	\label{eq3}
\end{equation}
The form of  Eq. \ref{eq3} is similar to the nonlinear Schrodinger equation, and the effective potential can be expressed under the metric  Eq. \ref{eq2} as ${V_{eff}}\left(h \right) \approx \left( {K/2} \right) - K\left( {3K/4-{k^2} } \right){h^2}$. Evidently, this is a harmonic oscillator potential, is the simplest potential field in quantum mechanics. In this paper, we would like to present a new mathematical method to obtain an analytical solution of the wave equation.

In the above research, it is demonstrated that the form of the Schrodinger equation is consistent with that of the Helmholtz equation under paraxial approximation, which implies that we can also introduce the method of path integration (PI) into the field of optics in curved space.

The classical diffraction integral is considered as the relationship between the initial input field   and the final output field as ${u_b}\left( {{h_b},{\rho _b}} \right) = \int {\exp \left( {ik{S_{cl}}} \right)} {u_a}\left( {{h_a},{\rho _a}} \right)d{h_a}$. ${S_{cl}}$ stands for a classical action, and here it is eikonal in the diffraction integral of light. 

In classical mechanics, the path of light is indeed unique. But, because of quantum randomness, photons can take any path, including geodesics, which has the highest probability. So, the diffraction integral formula for light is going to be rewritten for all possible paths $h\left( \rho  \right)$ from $a$ to $b$:
\begin{equation}
	\centering
	\begin{array}{l}
		u\left( {{h_b},{\rho _b}} \right) = \int {Q\left( {{h_a},{\rho _a};{h_b},{\rho _b}} \right)} u\left( {{h_a},{\rho _a}} \right)d{h_a},\\
		Q\left( {{h_a},{\rho _a};{h_b},{\rho _b}} \right)\!=\!\int\limits_a^b {\exp \left[ {ikS} \right]} Dh\left( \rho  \right)\!\\
		=\!\int\limits_a^b {\exp\!\left[ {ik\!\left( {\int\limits_{{\rho _a}}^{{\rho _b}} {dl} } \right)} \right]} Dh\left( \rho  \right).
	\end{array}
	\label{eq4}
\end{equation}
The convolution factor $Q\left( {a,b} \right)$ is called kernel in PI. In the diffraction integral, it's the exponential form of the eikonal. Here $Dh\left( \rho  \right)$ means a functional integral from $a$ to $b$. From a mechanical point of view, the optical path here actually corresponds to the action of Lagrangian integral. For the line element of CGCS,  the path integral kernel $Q\left( {a,b} \right)$ in  Eq. \ref{eq4} has an analytic solution under paraxial approximation:
\begin{equation}
	\centering
		\begin{aligned}
	Q\left( {a,b} \right) &= {\left( {\frac{{k\sqrt K }}{{2\pi i\sin \left( {\sqrt K \rho } \right)}}} \right)^{1/2}} \times\\
	&\exp \left( {ik\frac{{\sqrt K }}{{\sin \left( {\sqrt K \rho } \right)}}\left[ {\left( {{h_a}^2 + {h_b}^2} \right)\cos \left( {\sqrt K \rho } \right) - 2{h_a}{h_b}} \right]} \right),
	\label{eq5}
\end{aligned}
\end{equation}
where $\rho  = {\rho _b} - {\rho _a}$ . It is worth noting that this analytic solution is consistent with that obtained by the optical matrix and Collins formula in Ref. \cite{RN25}.  Eq. \ref{eq3} is a linear differential equation whose solutions satisfy the linear superposition property, so we can assume a special set of solutions called eigen-solutions ${u_n}\left( {h,\rho } \right)$ , which can be separated into the following form:
\begin{equation}\nonumber
	\centering
u\left( {h,\rho } \right) = f\left( \rho  \right)\phi \left( h \right)
\end{equation}
or
\begin{equation}\nonumber
		\centering
\frac{{df\left( \rho  \right)/d\rho }}{{f\left( \rho  \right)}} =  - \frac{i}{{2k}}\frac{{ - K{r^2}\left[ {{d^2}\phi \left( h \right)/d{h^2}} \right] + \phi \left( h \right)}}{{\phi \left( h \right)}}.
\end{equation}

Then we can construct $f\left( \rho  \right) = {f_0}\exp \left[ { - \left( {ip/\hbar } \right)\rho } \right]$ where ${f_0}$  is an arbitrary constant factor. Moreover, any transverse mode field $\phi \left( h \right)$ can be expressed as a linear combination of ${\phi _n}\left( h \right)$, which is not only normalized but also orthogonal. That is 
\begin{equation}
	\centering
	\phi \left( h \right) = \sum\limits_{n = 1}^\infty  {{a_n}} {\phi _n}\left( h \right)
	\label{eq6}.
\end{equation}
And the coefficients ${a_n} = \int\limits_{ - \infty }^\infty  {{\phi _n}^*\left( h \right)} {\phi }\left( h \right)dh$ . So, the light evolution from  $a$ to $b$ satisfies:
\begin{equation}
	\centering
	\begin{array}{l}
		u\left( {{h_b},\rho } \right) = \sum\limits_{n = 1}^\infty  {{a_n}{\phi _n}\left( {{h_b}} \right)\exp \left( { - i{p_n}\rho /\hbar } \right)} \\
		= \sum\limits_{n = 1}^\infty  {\left( {\int\limits_{ - \infty }^\infty  {{\phi _n}^*\left( {{h_a}} \right)} \phi \left( {{h_a}} \right)d{h_a}} \right){\phi _n}\left( {{h_b}} \right)\exp \left( { - i{p_n}\rho /\hbar } \right)} \\
		= \int\limits_{ - \infty }^\infty  {\sum\limits_{n = 1}^\infty  {{\phi _n}\left( {{h_b}} \right){\phi _n}^*\left( {{h_a}} \right)\exp \left( { - i{p_n}\rho /\hbar } \right)} } \phi \left( {{h_a}} \right)d{h_a}.
	\end{array}
	\label{eq7}
\end{equation}
Compared to  Eq. \ref{eq4}, we can obtain:
$Q\left( {{h_a},0;{h_b},\rho } \right) = \sum\limits_n {{e^{ - \frac{i}{\hbar }{p_n}\rho }}} {\phi _n}\left( {{h_b}} \right)\phi _n^*\left( {{h_a}} \right)$. 
Thus, combining with  Eq. \ref{eq5}, the eigenvalue ${p_n}$  is given by trace the Kernel $Q$ as  Eq. \ref{eq8}.
\begin{equation}
	\centering
	\begin{aligned}
		{\rm{tr}}\left( {Q\left( {a,b} \right)} \right) &= \int {Q\left( {{h_a},0;{h_a},\rho } \right)d{h_a}} \\
		&= \frac{{\exp \left( { - i\sqrt K \rho /2} \right)}}{{1 - \exp \left( { - i\sqrt K \rho } \right)}} = \sum\limits_{n = 0}^\infty  {\exp \left( { - i\left( {n + \frac{1}{2}} \right)\sqrt K \rho } \right)}.
	\end{aligned}
	\label{eq8}
\end{equation}
Particularly, for the eigenstate \!${u_n}\left( {h,\rho } \right)\!=\!{\phi _n}\!\left( h \right)\exp\!\left( { - i{p_n}\rho /\hbar } \right)$, the eigenvalue ${p_n}\!=\!\left( {n + \frac{1}{2}} \right)\hbar \sqrt K$ is the “quantized” momentum of the beam in CGCSs, which is proportional to the square root of the curvature. In flat space, this quantization disappears and the momentum of the beam can be measured continuously. Intriguingly, so far, we have not introduced any calculation based on quantum theory, only carried out the classical calculations in curved space, so this “quantized” momentum is different from the quantum coherent states whose energy levels are integer spaced. Furthermore, by physical interpretation, this momentum quantization can be derived by the standing wave condition of the interference cancellation. So, the circumference of the light path $2\pi r\!=\!2\pi /\sqrt K\!=\!(2n + 1)\lambda /2$, which is consistent to the “quantized” momentum condition. This peculiar quantized momentum generated by geometric effects is called geometric momentum and has been elucidated in mathematical physics \cite{RN26}.

Furthermore, due to the properties of Hermitian polynomials, we can expand the Kernel into a summarize form:
\begin{equation}\nonumber
	\centering
\begin{aligned}
	Q\left( {a,b} \right) &= {\left( {\frac{{k\sqrt K }}{\pi }} \right)^{1/2}}{e^{ - \frac{{k\sqrt K }}{2}\left( {{h_a}^2 + {h_b}^2} \right)}} \times \\
	&\sum\limits_{n = 0}^\infty  {\frac{1}{{{2^n}n!}}} {H_n}\left( {\sqrt {k\sqrt K } {h_a}} \right){H_n}\left( {\sqrt {k\sqrt K } {h_b}} \right){e^{ - i\left( {n + \frac{1}{2}} \right)\sqrt K \rho }}.
\end{aligned}
\end{equation}
\begin{figure}[htbp]
	\centering
	\includegraphics[width=0.7\linewidth]{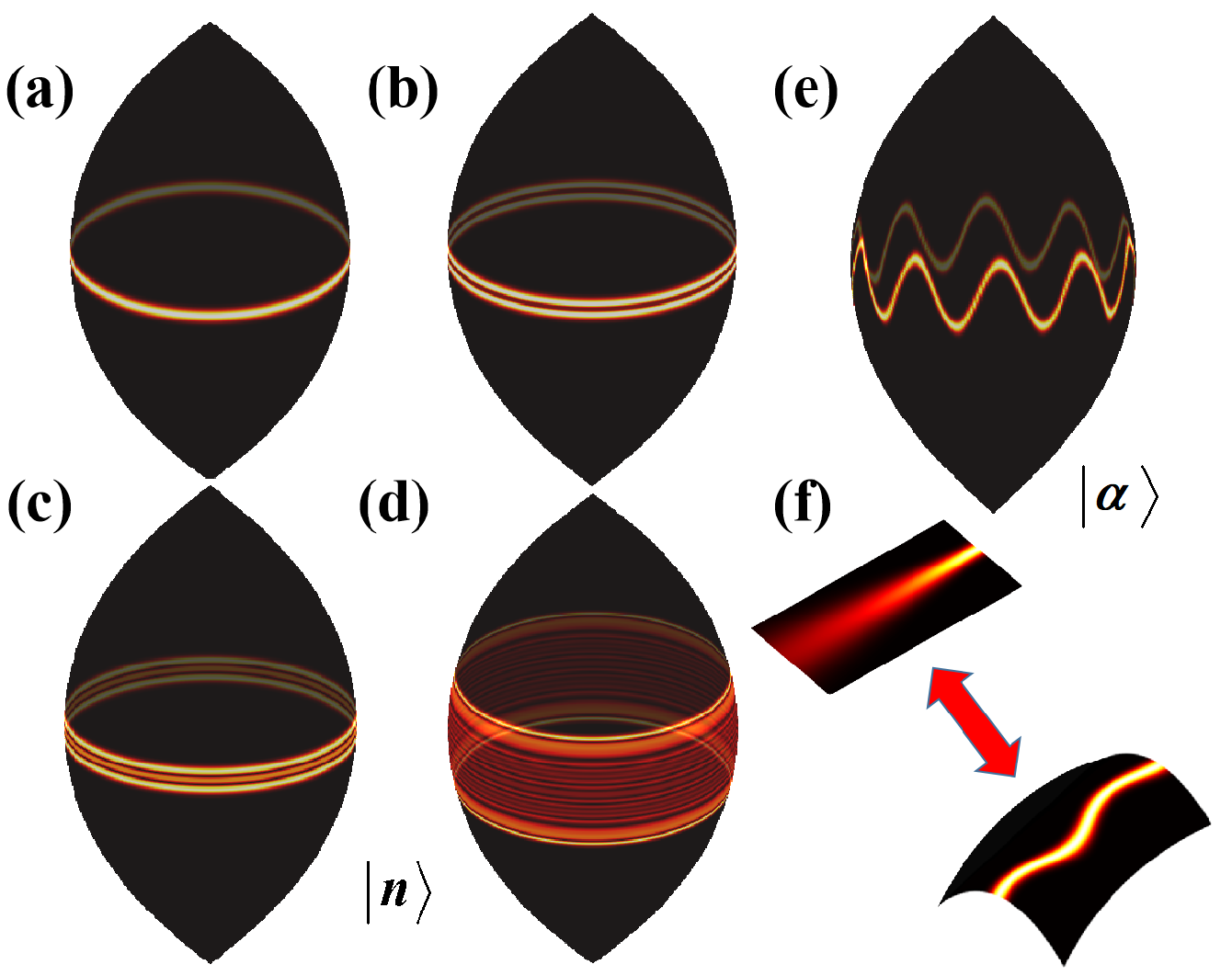}
	\caption{ Transmission of different optical fields on a curved surface. Graphs (a)$n=0$, (b)$n=1$, (c)$n=2$, (d)$n=50$ are the eigenstates $\left| n \right\rangle$ . Graph (e) depicts the evolution of a particular superposition of eigenstates. A comparison of Gaussian light in flat space and on curved surface is shown in graph (f).}
	\label{fig1}
\end{figure}

For the convenience of description, we use the language of quantum mechanics in the following, and treat the optical field as a photon state. By using the expansion of $Q$ , we can get the eigenstates based on the Dirac notation:
\begin{equation}
	\centering
		{\left| n \right\rangle _0}\!\equiv\!{\phi _n}\left( h \right)\!=\!{2^{ - n/2}}{(n!)^{ - 1/2}}{\left( {\frac{{k\sqrt K }}{\pi }} \right)^{1/4}}\! {H_n}\left( {\sqrt {k\sqrt K } h} \right){e^{ - \frac{{k\sqrt K }}{2}{h^2}}}.
\label{eq9}
\end{equation}

This description actually represents the eigen mode of the optical field. The superposition of eigenstates can be decomposed for any state $\left| a \right\rangle $, and its evolution can be written as $\left| a \right\rangle  = \sum\limits_n {{a_n}\left| n \right\rangle }  = \sum\limits_n {{a_n}{{\left| n \right\rangle }_0}{e^{ - i{p_n}\rho /\hbar }}}$. The momentum of any one of these states will also satisfy the superposition property.

Fig.\ref{fig1} (a)-(d) shows the propagation properties of different eigenstates, which can also be called number states because it represents the number of dark lines (the zero intensity of beam). On the other hand, we find that the number state $\left| n \right\rangle$ displays $n+1$ straight bright stripes on the curved surface. Although the stripes are not of equal width, the propagation of the beam shows a good non-diffraction property. As shown in Fig.\ref{fig1} (e), the evolution of this Gaussian light oscillates around the transmission axis, which foreshadows the construction of coherent states.

\section{“Classical” coherent state}
The evolution of light on the surface satisfies the paraxial wave equation  Eq. \ref{eq3}, just as the evolution of the quantum state fulfills the Schrodinger equation. We choose a special CGCS surface, which introduces a special effective potential - the harmonic oscillator potential. A photon may be thought of as a harmonic oscillator in certain ways, and the eigenstate of its annihilation operator (the falling operator) is a coherent state in the field of quantum optics. Similarly, in the previous section, we constructed the eigenstate of light transmission on a curved surface. Due to the superposition property of the optical field, we can also construct coherent states in a classical optical field. The coherent states $\left| \alpha  \right\rangle $ are defined by using some superposition of the number states $\left| n \right\rangle$:
\begin{equation}
	\centering
	\left| \alpha  \right\rangle  = A{e^{ - {{\left| \alpha  \right|}^2}/2}}\sum\limits_n {\frac{{{\alpha ^n}}}{{\sqrt {n!} }}} \left| n \right\rangle ,
	\label{eq10}
\end{equation}
where parameter ${\left| \alpha  \right|^2} = \bar n$. By substituting   Eq. \ref{eq9}, we can obtain its representation in coordinates:
\begin{equation}
	\centering
	{u_\alpha }\left( {h,\rho } \right)\!=\!\left\langle {h}
	\mathrel{\left | {\vphantom {h \alpha }}
		\right. \kern-\nulldelimiterspace}
	{\alpha } \right\rangle \!=\!\frac{{A\exp \left\{ { - \frac{{k\sqrt K }}{2}\left[ {{h^2} - 2{h_0}{e^{ - i\sqrt K \rho }} + \frac{{{h_0}^2}}{2}\left( {1 + {e^{ - 2i\sqrt K \rho }}} \right)} \right]} \right\}}}{{\sqrt {\cos \left( {\sqrt K \rho } \right) - i\sin \left( {\sqrt K \rho } \right)} }}.
	\label{eq11}
\end{equation}
Obviously, this is an off-axis Gaussian beam of an initial condition: ${u_\alpha }\left( {h,0} \right) = A\exp \left[ { - \sqrt K k{{\left( {h - {h_0}} \right)}^2}/2} \right]$. The initial center of the beam is ${h_0} = \sqrt {\frac{2}{{k\sqrt K }}} \alpha$  and the spot size is a particular value ${1 \mathord{\left/{\vphantom {1 {\sqrt {\sqrt K k} }}} \right.\kern-\nulldelimiterspace} {\sqrt {\sqrt K k} }}$. This expression of Gaussian beam with special spot size is a so-called function of coherent state.

Next, without loss of generality, the beams of arbitrary spot size need to be studied. In general, we can set the initial Gaussian field as $u\left( {h,0} \right)\!=\!A\exp \left[ { - {{\left( {h - {h_0}} \right)}^2}/\left( {2{\sigma _0}^2} \right)} \right]$, ${\sigma _0}$ is the initial spot, ${z_r} = k{\sigma _0}^2/2$ is the Rayleigh distance of the beam. According to the diffraction integrate formula, we can obtain the following expression of the transmission field.
\begin{equation}
	\centering
	\begin{aligned}
		&{u_\sigma }\left( {h,\rho } \right) = \frac{{A{{\left( {k{\sigma _0}^2\sqrt K } \right)}^{1/2}}}}{{\sqrt {2k{\sigma _0}^2\sqrt K \cos \left( {\sqrt K \rho } \right) + i\sin \left( {\sqrt K \rho } \right)} }} \times \\
		&\exp \left\{ {\frac{{k\sqrt K \left[ { - \left( {{h_0}^2 + {h^2}} \right)\cos \left( {\sqrt K \rho } \right) + 2h\left[ {{h_0} - ihk{\sigma _0}^2\sqrt K \sin \left( {\sqrt K \rho } \right)} \right]} \right]}}{{2k{\sigma _0}^2\sqrt K \cos \left( {\sqrt K \rho } \right) + i\sin \left( {\sqrt K \rho } \right)}}} \right\}.
	\end{aligned}
	\label{eq12}
\end{equation}
\begin{figure}[htbp]
	\centering
	\includegraphics[width=0.85\linewidth]{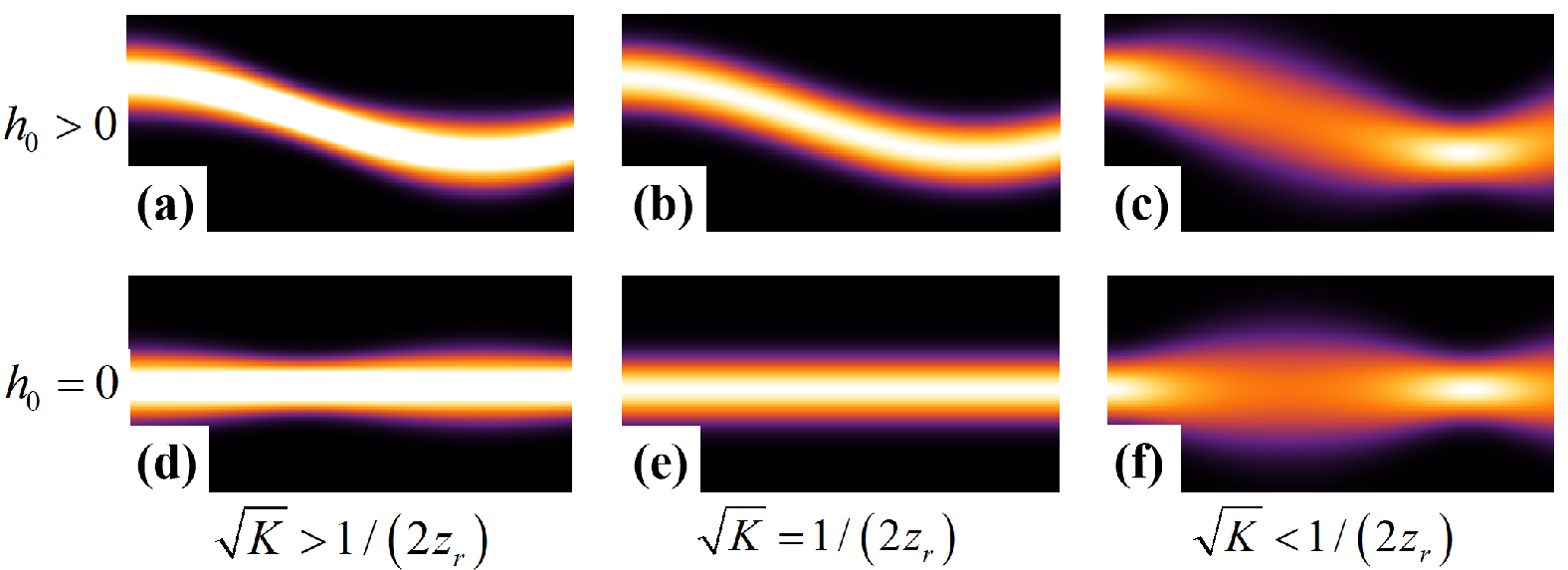}
	\caption{ Propagation of Gaussian beams with (the first row) and without (the second row) misalignment on surfaces of different curvatures. Graphs (a) and (d) show the beam focusing and then diverging when $\sqrt K  > 1/\left( {2{z_r}} \right)$. Graphs (b) and (e) show the non-diffraction propagation of beam when $\sqrt K  = 1/\left( {2{z_r}} \right)$ . Graphs (c) and (f) show the beam diverging and then focusing when $\sqrt K  < 1/\left( {2{z_r}} \right)$.}
	\label{fig2}
\end{figure}

Fig. \ref{fig2} shows the transmission of Gaussian beam with the same initial spot on the different CGCS. We can explore that the diffraction properties of Gaussian beam contend with the focusing effect of positive curvature in space. It is the initial beam size and the space curvature that determine the divergence or focus of the light beam. Intuitively, as shown in Fig. \ref{fig2} (b) and (e), the size of the beam spot does not change under a special matching condition $\sqrt K  = 1/\left( {2{z_r}} \right)$. This good non-diffraction property can be used as a simulation of oscillatory Gaussian wave packets under the quadrature representation of the coherent state. 

\begin{figure}[htbp]
	\centering
	\includegraphics[width=0.75\linewidth]{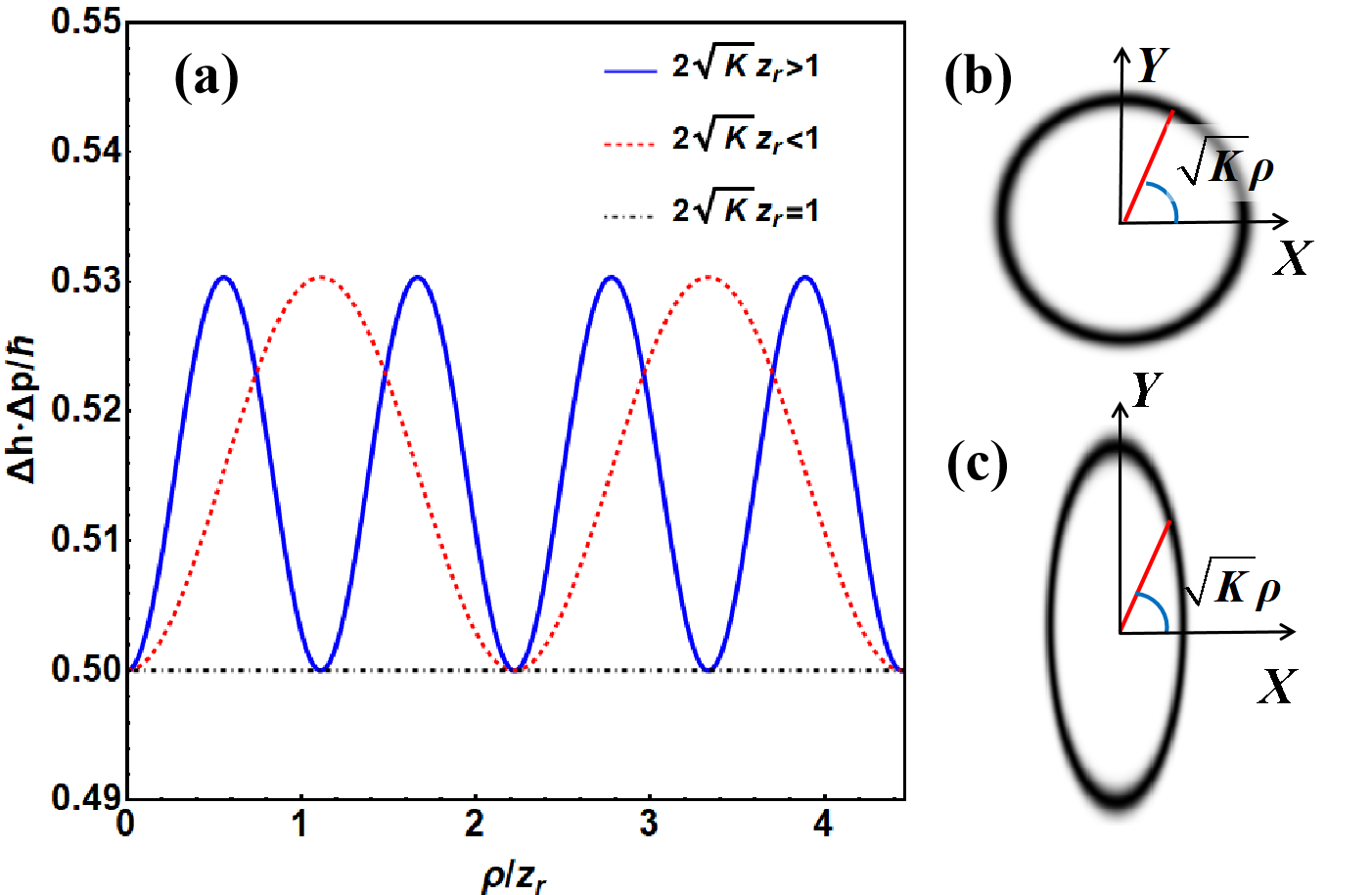}
	\caption{Demonstration of the momentum and position uncertainty. (a) Evolution of the uncertainty product under three different conditions. Orthogonal phase space diagram of (b) "classical" coherent state and (c) "classical" squeezed state.}
	\label{fig3}
\end{figure}

The coherent state is also known as the minimal uncertainty state, here our "classical" coherent state does as well. Generally, for Gaussian beams of arbitrary spot size, we can obtain the evolution of beam’s position and momentum:
\begin{equation}
	\centering
	\begin{aligned}
		&\left\langle h \right\rangle  = \frac{{\int {{u_\sigma }h{u_\sigma }^*dh} }}{{\int {{u_\sigma }{u_\sigma }^*dh} }} = {h_0}\cos \left( {\sqrt K \rho } \right),\\
		&\left\langle {{p_h}} \right\rangle  = \frac{{\int {{\eta _\sigma }{p_h}{\eta _\sigma }^*dh} }}{{\int {{\eta _\sigma }{\eta _\sigma }^*dh} }} = \hbar k\sqrt K {h_0}\sin \left( {\sqrt K \rho } \right).
	\end{aligned}
	\label{eq13}
\end{equation}
${\eta _\sigma }\left( {{p_h}} \right) = \int {{u_\sigma }} {e^{ - i{p_h}h}}dh$ is the Fourier transform of ${u_\sigma }$ . According to  Eq. \ref{eq13}, we can find that the transmission of the light beam along  axis can be regarded as the rotation in the position-momentum ($h - {p_h}$) space, which coincides with our previous discussion that CGCS can be regarded as a fractional Fourier transform system \cite{RN23,RN25}. Furthermore, the coordinate uncertainty (spot size) and momentum uncertainty can be obtained by the second moment:
\begin{equation}
	\centering
	\begin{aligned}
		&{\left( {\Delta h} \right)^2} = \frac{{K{k^2}{\sigma _0}^2 + \left( {1/{\sigma _0}^2} \right) + \left( {K{k^2}{\sigma _0}^2 - 1/{\sigma _0}^2} \right)\cos \left( {2\sqrt K \rho } \right)}}{{4K{k^2}}},\\
		&{\left( {\Delta {p_h}} \right)^2} = \frac{{K{k^2}{\sigma _0}^2 + \left( {1/{\sigma _0}^2} \right) - \left( {K{k^2}{\sigma _0}^2 - 1/{\sigma _0}^2} \right)\cos \left( {2\sqrt K \rho } \right)}}{4}{\hbar ^2}.
	\end{aligned}
	\label{eq14}
\end{equation}
Here, the notation ${\left( {\Delta \mathord{\buildrel{\lower3pt\hbox{$\scriptscriptstyle\frown$}} \over O} } \right)^2} = \left\langle {{{\mathord{\buildrel{\lower3pt\hbox{$\scriptscriptstyle\frown$}} \over O} }^2}} \right\rangle  - {\left\langle {\mathord{\buildrel{\lower3pt\hbox{$\scriptscriptstyle\frown$}} \over O} } \right\rangle ^2}$. From \ref{eq14}, parameters $X = h\left( {k{\sigma _0}\sqrt K } \right)$ and $Y = {p_h}{\left( {k{\sigma _0}\sqrt K } \right)^{ - 1}}/\hbar $  satisfy an elliptic equation and the uncertainty principle can be given as follows:
\begin{equation}
	\centering
	{\left( {\Delta h} \right)^2}{\left( {\Delta {p_h}} \right)^2} = \frac{1}{4}{\hbar ^2} + {\left( {\frac{{\left( {K{k^2}{\sigma _0}^2 - 1/{\sigma _0}^2} \right)\left( {\sin 2\sqrt K \rho } \right)}}{{4\sqrt K k}}} \right)^2}{\hbar ^2} \ge \frac{1}{4}{\hbar ^2}.
	\label{eq15}
\end{equation}
The equal sign in “$\ge$ ” symbol is always true if and only if $\sqrt K  = 1/\left( {2{z_r}} \right)$ , which is also called a non-diffraction condition. Besides, the uncertainty product reaches the minimum value $\Delta h\Delta {p_h} = \hbar /2$. To better illustrate this point, we can introduce two dimensionless orthogonal parameters $X = h\left( {k{\sigma _0}\sqrt K } \right)$ and $Y = {p_h}{\left( {k{\sigma _0}\sqrt K } \right)^{ - 1}}/\hbar$ as shown in Fig. \ref{fig3} (b) and (c). In the "classical" coherent state case, ${\left( {\Delta X} \right)^2} = {\left( {\Delta Y} \right)^2} = 1/2$.

In addition, in the process of beam evolution, the uncertainty product will periodically reach its minimum at $\sin \left( {2\sqrt K \rho } \right) = 0$, as shown in Fig. \ref{fig3} (a). At these special points, we find that although the product of two uncertainties still satisfies the Heisenberg relation, the uncertainty of one of the momentum or position is compressed as  Eq. \ref{eq16}. This is similar to the "squeezed state" in QO as shown in Fig. \ref{fig3} (c).
\begin{equation}
	\centering
	\begin{aligned}
		&\Delta {X_s} = \Delta {X_\alpha } \cdot \gamma, \\
		&\Delta {Y_s} = \Delta {Y_\alpha }/\gamma,
	\end{aligned}
	\label{eq16}
\end{equation}
where $\gamma  = 2\sqrt K {z _r}$  is the squeeze factor. Evidently, $\gamma  = 1$ when the non-diffraction condition is satisfied, which goes back to the "classical" coherent state.

Coherent states and compressed states are quantum optics concepts, and in fact, they also directly undergo the similar reciprocal transformation process. To some extent, this classical light field exhibits the quantum properties of a first quantization, but it is clear that we have difficulty in dealing with a second quantization.
We give a new physical understanding of light transmission on a curved surface, and its evolution is similar to that of coherent states, which provides ideas for us to conduct quantum optical simulation experiments from another perspective.

\section{Conclusion}
We deal with how classical optical fields, in particular Gaussian beams, behave in a curved surface and how a "classical" coherent states approach may aid in physical understanding of the light transmission on curved surfaces. The idea is to use some quantum techniques, such as the path integral, in this classical setting to obtain a new set of solutions that don't occur in flat space-time. Under a positive curvature, the CGCS metric can restrict the divergence of Gaussian beams, and both momentum and position uncertainty fluctuate periodically. Surprisingly, we find that the transmission of a basic classical optical field on curved surface generates one simulated coherent state and naturally yields quantized momentum along the transmission direction in this research. We demonstrate the feasibility of creating such quantum states from spatial curvature theoretically.  

According to Einstein, gravity can bend spacetime, as this cross section of FRW space-time can be proved to correspond to a curved surface of constant Gaussian curvature and its shape can be
approximated invariant in the short-term transmission of light. Therefore, the research on this surface can theoretically demonstrate the light transmission in gravitational field, and can be used to calculate the transmission mode of astronomical information. In addition, we think the study is not only of theoretical interest, but also can construct some special nanostructures with novel features \cite{RN27,RN28,RN29} due to the localization of light on the surface, which can be used to the classical simulation of quantum coherent states.

%
%
%
%
\section*{Reference}





\end{document}